\documentclass[a4paper]{article}
\newcommand{\be} {\begin{equation}}
\newcommand{\ee} {\end{equation}}
\newcommand{\bdm} {\begin{displaymath}}
\newcommand{\edm} {\end{displaymath}}
\newcommand{\bc} {\begin{center}}
\newcommand{\ec} {\end{center}}
\newcommand{\beqa} {\begin{eqnarray}}
\newcommand{\eeqa} {\end{eqnarray}}

\newcommand{\bfig} {\begin{figure}}
\newcommand{\efig} {\end{figure}}
\newcommand{\btab} {\begin{tabular}}
\newcommand{\etab} {\end{tabular}}
\newcommand{\hl} {\hline}
\usepackage{epsfig}
\usepackage{amssymb}
\usepackage{amsmath}
\begin{document}
\bc
{\bf\Large SCALAR MESON PHOTOPRODUCTION}
\ec
\bigskip
\bc
{\large A Donnachie}\\
{\large School of Physics and Astronomy, University of Manchester}\\
{\large Manchester M13 9PL, England}
\ec
\medskip
\bc
{\large Yu S Kalashnikova}\\
{\large ITEP}\\
{\large 117259 Moscow, Russia}
\ec
\medskip
\begin{abstract}
\noindent
The scalar mesons $f_0(1370)$, $f_0(1500)$ and $f_0(1710)$ are of interest 
as there is as yet no consensus of their status, or indeed of the existence 
of the $f_0(1370)$. Radiative decays to $\rho$ and $\omega$ have been shown
to provide effective probes of their structure and to discriminate among 
models. Scalar-meson photoproduction is proposed as an alternative and 
it is shown that it represents a feasible approach.
\end{abstract}
{\large
\section{Introduction}

The fundamental structure of the light scalar mesons is still a subject of 
debate. The $a_0(1450)$ and the $K^*_0(1430)$ are generally regarded as the 
$u\bar{d}$ and $u\bar{s}$ members of the same SU(3) flavour nonet, to which 
the $f_0(1370)$ can be attached as the $(u\bar{u}+d\bar{d})$ 
member\cite{PDG06}. There then remain two possibilities for the ninth member 
of the nonet, the $f_0(1500)$ and the $f_0(1710)$. In this picture, it is 
usually assumed that the surplus of isoscalar scalars in the 1300 to 1700 MeV 
mass region can be attributed to the presence of a scalar glueball. This 
assumption has been supported in the past by calculations in quenched LQCD, 
which predict a scalar glueball in this mass range\cite{MP99}. 
The three physical states are then viewed as mixed $q\bar{q}$ and gluonium 
states, although there is not agreement in detail about the 
mixing\cite{LW00,CK01}. However calculations in unquenched LQCD\cite{LQCD} 
suggest that there is a sizeable contribution from glueball interpolating 
operators  to the states around or below 1 GeV, casting some 
doubt on the mixing models. Further, it has been argued that the $f_0(1370)$
may not exist\cite{KZ07,Ochs06}. This is strongly contested by Bugg\cite{Bugg}.
If the $f_0(1370)$ does not exist the lowest scalar nonet can be taken to 
comprise the $a_0(980)$, the $f_0(980)$, the $f_0(1500)$ and the $K_0^*(1430)$,
the $f_0(980)$ and the $f_0(1500)$ being mixed such that the former is close 
to a singlet and the latter close to an octet. The lightest scalar glueball is 
then considered to be a broad object extending from 400 MeV to about 1700 MeV. 
So a variety of interpretations is possible.

Radiative transitions offer a particularly powerful means of probing the
structure of hadrons as the coupling to the charges and spins of the
constituents reveals detailed information about wave functions and can 
discriminate among models. In the case of the $f_0(1370)$, $f_0(1500)$ 
and $f_0(1710)$ being considered as mixed $n\bar n$, $s\bar s$ and glueball 
states their radiative decays to a vector meson, $S \to V\gamma$, are strongly 
affected by the degree of mixing between the basis $q\bar{q}$ states and the 
glueball\cite{CDK03}. Three different mixing scenarios have been proposed:
the bare glueball is lighter than the bare $n\bar{n}$ state\cite{CK01}; its 
mass lies between the bare $n\bar{n}$ state and the bare $s\bar{s}$ 
state\cite{CK01}; or it is heavier than the bare $s\bar{s}$ state\cite{LW00}.
We label these three possibilites L, M, H respectively. Assuming that the 
$q\bar{q}$ basis of the $f_0(1370)$, $f_0(1500)$ and $f_0(1710)$ is in the 
$1^3P_0$ nonet, the discrimination among the different mixing scenarios is 
strong\cite{CDK03}. Preliminary results on the implications of this particular 
scenario for photoproduction are presented here. 
 
Photoproduction of the scalar mesons at medium energy provides an alternative 
to direct observation of the radiative decays. It is this possibility that we 
explore here and show that it is viable. The dominant mechanism is Reggeised
$\rho$ and $\omega$ exchange, both of which are well understood in pion
photoproduction\cite{GLV97}. The energy  must be sufficiently high for
the Regge approach to be applicable but not too high as the cross section 
decreases approximately as $s^{-1}$. In practice this means approximately
5 to 10 GeV photon energy. In addition to photoproduction on protons 
we consider coherent photoproduction on $^4$He, encouraged in this by a 
recently-approved experiment at Jefferson Laboratory\cite{JLab}. 

\section{The Model}

The differential cross section is given by
\be
\frac{d\sigma}{dt}=\frac{|M(s,t)|^2}{64\pi|{\bf p}|^2s}.
\label{dsigma}
\ee
For the exchange of a single vector meson
\beqa
|M(s,t)|^2 &=& \textstyle{\frac{1}{2}}A^2(s,t)(s(t-t_1)(t-t_2)+
\textstyle{\frac{1}{2}}st(t-m_S^2)^2) \nonumber\\
&&+A(s,t)B(s,t)m_ps(t-t_1)(t-t_2) \nonumber\\
&&+\textstyle{\frac{1}{8}}B(s,t)^2s(4m_p^2-t)(t-t_1)(t-t_2).
\label{msquare}
\eeqa
where $t_1$ and $t_2$ are the kinematical boundaries
\beqa
t_{1,2} &=& \frac{1}{2s}\Big(-(m_p^2-s)^2+m_S^2(m_p^2+s)\nonumber\\
&&\pm (m_p^2-s)\surd((m_p^2-s)^2-2m_s^2(m_p^2+s)+m_S^4)\Big),
\label{t12}
\eeqa
and 
\be
A(s,t)=\frac{g_S(g_V-2 m_p g_T)}{m_V^2-t},~~~~B(s,t)=-\frac{2g_Sg_T}{m_V^2-t}.
\label{abdef}
\ee
In (\ref{abdef}), $g_V$ and $g_T$ are the $VNN$ vector and tensor couplings, 
$g_S$ is the $\gamma V N$ coupling. The $\omega N N$ couplings are rather well 
defined\cite{RHE87}, with $13.8 < g_V^{\omega} < 15.8$ and $g_T^\omega 
\approx 0$. We have used $g_V^{\omega} = 15$ and $g_T^\omega =0$ as this gives 
a good description of $\pi^0$ photoproduction\cite{GLV97}. The $\rho N N$ 
couplings are not so well defined, with two extremes: strong 
coupling\cite{RHE87} or weak coupling\cite{OO78}. We are again 
guided by 
pion photoproduction\cite{GLV97} and choose the strong coupling solution with 
$g_V^{\rho} = 3.4$, $g_T^{\rho}= 11$ GeV$^{-1}$. The $S V \gamma$ coupling, 
$g_S$, can be obtained from the radiative decay width through\cite{KKNHN06}
\be
\Gamma(S \to \gamma V) = g_S^2\frac{m_S^3}{32\pi}\Bigg(1-\frac{m_V^2}{m_S^2}
\Bigg)^3.
\label{width}
\ee
Obviously in practice the amplitudes for $\rho$ and $\omega$ exchange are
added coherently.

The standard prescription for Reggeising the Feynman propagators in 
(\ref{msquare}), assuming a linear Regge trajectory $\alpha_V(t)= \alpha_{V0}+
\alpha^\prime_V t$, is to make the replacement
\be
\frac{1}{t-m_V^2} \to \Big(\frac{s}{s_0}\Big)^{\alpha_V(t)-1}
\frac{\pi\alpha^\prime_V}{\sin(\pi\alpha_V(t))}
\frac{-1+e^{-i\pi\alpha_V(t)}}{2}\frac{1}{\Gamma(\alpha_V(t))}.
\label{regge}
\ee
This simple prescription automatically includes the zero observed at 
$t \approx -0.6$ GeV$^2$ in both $\rho$ and $\omega$ exchange and provides
a satisfactory description of the $\rho$ and $\omega$ exchange contributions 
to pion photoproduction\cite{GLV97}. 

\bfig[t]
\bc
\begin{minipage}{70mm}
\epsfxsize70mm
\epsffile{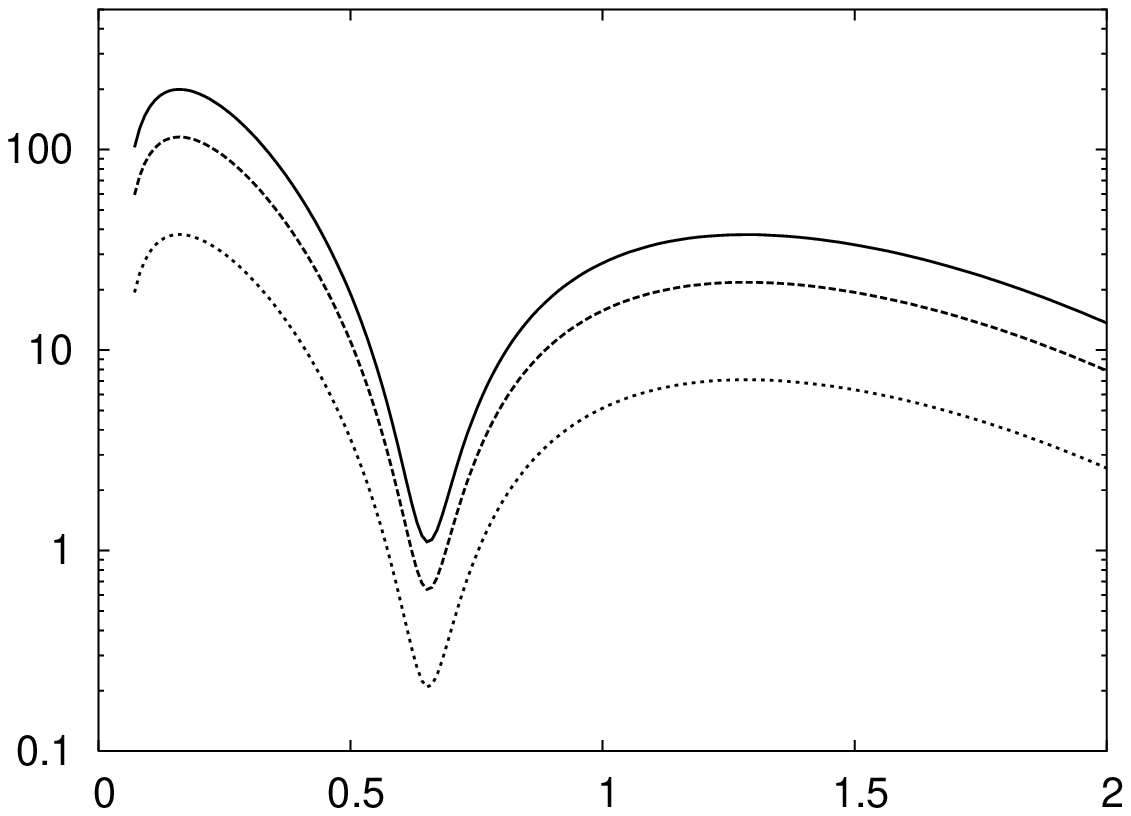}
\begin{picture}(0,0)
\setlength{\unitlength}{1mm}
\put(-6,38){{\tiny $d\sigma/dt$}}
\put(35,45){{\tiny $f_0(1500)$}}
\put(-9,34){{\tiny (nb GeV$^{-2}$)}}
\put(40,1){{\tiny $-t$ (GeV$^2$)}}
\end{picture}
\end{minipage}
\caption{Differential photoproduction cross section on hydrogen for 
$f_0(1500)$ at $E_\gamma =5$ GeV. The glueball masses are L (solid), 
M (dashed) and H (dotted).}
\ec
\label{dsigdt}
\efig

For photoproduction on $^4$He we assume that the cross section is given by
\be
\frac{d\sigma(\gamma N \to f_0 {\rm He})}{dt} = \frac{d\sigma(\gamma N \to 
f_0 N)}{dt}\Big(4F_{\rm He}(t)\Big)^2,
\label{helium1}
\ee
where $F_{\rm He}(t)$ is the helium form factor\cite{MS91}, 
$F_{\rm He}(t) \approx e^{9 t}$.
The justification for the assumption (\ref{helium1}) is the low level of 
nuclear shadowing observed on $^4$He at the energies with which we are 
concerned, for both pion and photon total cross sections\cite{GS78}. 

We assume non-degenerate $\rho$ and $\omega$ trajectories
\be
\alpha_\rho = 0.55 + 0.8 t,~~~~\alpha_\omega = 0.44 + 0.9 t.
\label{traj}
\ee  

\section{Cross Sections}

The differential cross sections have the structure expected, that is vanishing
in the forward direction due to the helicity flip at the photon-scalar vertex
and having a deep dip at $ -t \approx 0.6$ GeV$^2$ due to the zeroes in the 
exchange amplitudes in (\ref{regge}). It does not go to zero in the dip 
because of the non-degenerate trajectories (\ref{traj}). This is illustrated
for $f_0(1500)$ photoproduction at $E_\gamma = 5$ GeV in figure 1. The 
integrated cross sections for photoproduction of the scalars on protons and 
$^4$He at $E_\gamma$ = 5 GeV are given in table 1 for light (L), medium (M) 
and heavy (H) glueball masses. In the case of $^4$He the integration over 
$d\sigma/dt$ is for $|t| > 0.1$ GeV$^2$ due to the experimental requirement 
that $|t| \gtrsim 0.1$ GeV$^2$ for the recoiling helium to be detected. The 
cross sections for photoproduction on protons at higher energies are similar 
in shape, but the magnitude decreases with energy at the rate expected from 
(\ref{regge}). For example the cross sections at $E_\gamma = 10$ GeV are about 
half those in table 1. However the cross sections for photoproduction on 
$^4$He do not decrease, and for the $f_0(1500)$ and $f_0(1710)$ actually 
increase. This is due the combined effect of the $^4$He form factor enhancing 
the contribution from small $t$ and the maximum of the differential cross 
section on protons moving to smaller $|t|$ with increasing energy.

The reasons for the cross sections for scalar photoproduction on $^4$He being 
very much smaller than those for scalar photoproduction on protons are (i) 
switching off $\rho$ exchange for photoproduction on protons reduces 
the cross section by a factor of about 16, cancelling the factor 16 from 
coherent production (ii) the helium form factor suppresses the cross section 
except at very small $t$ (iii) there is the experimental requirement that 
$|t| \gtrsim 0.1$ GeV$^2$ for the recoiling helium to be detected.

\begin{table}
\bc
\caption{\it Integrated photoproduction cross sections in nanobarns on protons 
and $^4$He at $E_\gamma = 5$ GeV for the three different mixing scenarios: 
light glueball (L), medium-weight glueball (M) and heavy glueball (H).}
\vskip2.5mm
\btab{|c|c|c|c|c|c|c|}
\hl
              & \multicolumn{3}{c|}{proton}   
              & \multicolumn{3}{c|}{$^4$He}         \\
\hl
Scalar & L & M & H & L & M & H \\
\hl
$f_0(1370)$ & 27.1 & 68.6 & 94.2 & 0.64 & 1.63 & 2.23 \\
$f_0(1500)$ & 89.9 & 52.1 & 17.0 & 1.55 & 0.90 & 0.29 \\
$f_0(1710)$ &  0.7 &  1.6 & 11.8 & 0.001 & 0.002 & 0.016 \\
\hl
\etab
\ec
\label{sigma}
\end{table}

The cross sections in table 1 reflect directly the radiative decay widths and, 
if it were practical, ratios of cross sections $f_0(1370):f_0(1500):f_0(1710)$ 
would give an immediate result and ``weigh'' the glueball. In practice there 
are several problems in realising this ideal scenario. It is unlikely that the 
decay modes of the scalars with charged particles can be considered because of 
the very much larger cross sections in $\pi^+\pi^-$, $K^+K^-$, $2\pi^+2\pi^-$ 
and $\pi^+\pi^-2\pi^0$ from vector-meson production. The contribution from 
vector mesons can be eliminated by considering only the all-neutral channels, 
that is the $\pi^0\pi^0$, $\eta^0\eta^0$ and $4\pi^0$ decays of the 
$f_0(1370)$, $f_0(1500)$ and $f_0(1710)$. A further difficulty is the 
uncertainty in the branching fractions of the $f_0(1370)$ and $f_0(1710)$, 
particularly the former\cite{PDG06,KZ07}, and the small cross section for the 
$f_0(1710)$. 

\begin{table}
\bc
\caption{\it Branching fractions in percent for the $f_0(1500)$ from the 
PDG\cite{PDG06}, the WA102 experiment\cite{WA102} from the analysis of Close 
and Kirk\cite{CK01} (CK) and the Crystal Barrel experiment\cite{CB01} (CB).}
\vskip2.5mm
\btab{|l|c|c|c|}
\hl
Channel & PDG & WA102/CK & CB \\
\hl
$\pi\pi$          & $34.9 \pm 2.3$ & $33.7 \pm  3.4$ & $33.9 \pm  3.7$ \\
$\eta\eta$        & $ 5.1 \pm 0.9$ & $ 6.1 \pm  0.1$ & $ 2.6 \pm  0.3$ \\
$\eta\eta^\prime$ & $ 1.9 \pm 0.8$ & $ 3.2 \pm  0.7$ & $ 2.2 \pm  0.1$ \\
$K\bar{K}$        & $ 8.6 \pm 0.1$ & $10.7 \pm  2.4$ & $ 6.2 \pm  0.5$ \\
$4\pi$            & $49.5 \pm 3.3$ & $46.3 \pm  8.5$ & $55.1 \pm 16.9$ \\
\hl
\etab
\ec
\label{bf}
\end{table}

In contrast the cross sections for photoproduction of the $f_0(1500)$ on
protons are reasonable and the branching fractions are well defined. This 
is demonstrated in table 2 in which the branching fractions, in percent, are
given from the PDG\cite{PDG06}, the WA102 experiment\cite{WA102} as obtained in
the analysis of Close and Kirk\cite{CK01} and the Crystal Barrel 
experiment\cite{CB01}. Thus photoproduction of the $f_0(1500)$ on protons 
is the benchmark experiment and the obvious all-neutral channel is 
$\pi^0\pi^0$, although it should be recalled that the $\pi\pi$ branching 
fraction shown in table 2 has to be divided by a factor of three. From 
\hbox{table 1} 
we see that the ratio ${\rm L}:{\rm M} = 1.7:1$ and the ratio 
${\rm L}:{\rm H} = 5.3:1$. The latter is certainly appreciably larger than 
the uncertainties in the model.

Photoproduction of the $f_0(1370)$ can help resolve the ambiguities discussed
in the Introduction. Quite apart from the possibility that it does not
exist\cite{KZ07,Ochs06}, there is considerable variance in the branching 
fractions. In the analysis of Close and Kirk\cite{CK01} $4\pi$ is the dominant
decay mode, with a branching fraction of about $95\%$, and the $\pi\pi$
branching fraction is very small, $(2.7 \pm 1.2)\%$. This pattern is
replicated by Crystal Barrel\cite{CB01}, with a $4\pi$ branching fraction
of about $85\%$ and a $\pi\pi$ branching fraction of $(7.9 \pm 2.9)\%$.
In direct contrast, $\pi\pi$ is the dominant decay mode in the analysis 
of Bugg\cite{Bugg} and $4\pi$ is small. At resonance the ratio $2\pi:4\pi$
is given as $6:1$. However for the $f_0(1500)$ the $2\pi:4\pi$ ratio is 
$0.9:1$ so is not incompatible with table 2.  

\bfig[t]
\bc
\begin{minipage}{60mm}
\epsfxsize60mm
\epsffile{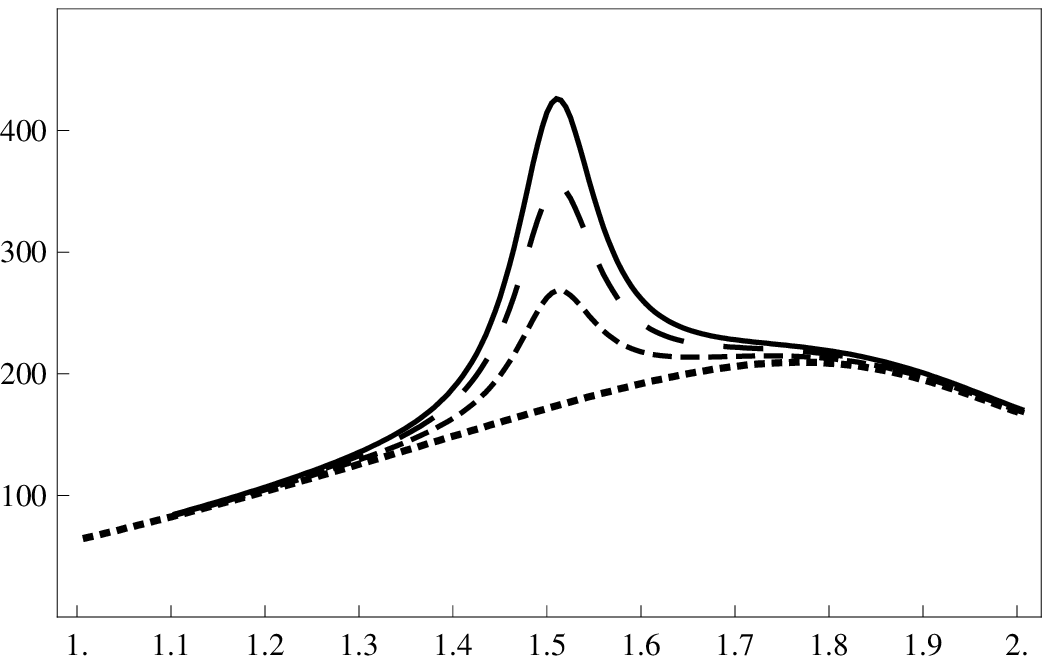}
\begin{picture}(0,0)
\setlength{\unitlength}{1mm}
\put(-10,34){{\tiny $d\sigma/dM$}}
\put(10,35){{\tiny $\pi^0\pi^0$}}
\put(-12,30){{\tiny (nb GeV$^{-1}$)}}
\put(40,1){{\tiny $M$ (GeV)}}
\end{picture}
\end{minipage}
\ec
\label{back}
\caption{Continuum $\pi^0\pi^0$ background (dotted) and combined with 
$f_0(1500)$ at $E_\gamma = 5$ GeV (solid (L), large dashed (M), small dashed 
(H)) for constructive interference.}
\efig

Of course the scalars are not produced in isolation. For example in the 
$\pi^0\pi^0$ channel there is a continuum background arising from the process 
$\gamma \to \pi^0 \omega (\rho)$ with subsequent rescattering of the 
$\omega (\rho)$ on the proton by $\rho (\omega)$ exchange to give the second 
$\pi^0$. The new ingredients 
here are the $\gamma\pi^0\omega(\rho)$ and $\omega\pi^0\rho$ couplings,
which can be estimated from\cite{Bramon92}. The $\rho(\omega)$ exchange is 
Reggeised as before. Figure 2 shows the result of this calculation together 
with the result of constructive interference with the $f_0(1500)$ signal.

}
\end{document}